\documentclass[%
  twoside,
  reprint,
  amsmath,amssymb,
  aps,
  prl,
  nofootinbib,
  a4paper,
]{revtex4-1}

\usepackage{graphicx}
\usepackage{dcolumn}
\usepackage{bm}
\usepackage{subfigure}
\usepackage[usenames,dvipsnames]{xcolor}
\usepackage{relsize} 

\usepackage[
text={7.3in,10in},centering,
total={6.5in,8.75in}, top=1.0in, left=0.6in, includefoot,
]{geometry}

\usepackage[
  pdfstartview={XYZ null null 1},
  bookmarks=true,
  pdfpagemode=UseNone,
  colorlinks,
  linkcolor=blue,
  urlcolor=blue,
  citecolor=blue,
  plainpages=false,
  pdfpagelabels,
  final,
  breaklinks=true
]{hyperref}
\hypersetup{
pdftitle={Momentum transfers in correlation-assisted tunnelling}, 
pdfauthor={E Pisanty and M Ivanov}
}

\renewcommand{\d}{\ensuremath{\textrm{d}}}
\renewcommand{\Re}{\operatorname{Re}}
\renewcommand{\Im}{\operatorname{Im}}
\newcommand{\bra}[1]{\ensuremath{\left\langle#1\right|}}
\newcommand{\ket}[1]{\ensuremath{\left|#1\right\rangle}}
\newcommand{\bracket}[2]{\ensuremath{\left\langle#1 \vphantom{#2}\middle|  #2 \vphantom{#1}\right\rangle}}

\newcommand{\vb}[1]{\mathbf{#1}}
  \newcommand{\vbr}{\vb{r}}
  \newcommand{\vbk}{\vb{k}}
  \newcommand{\vbp}{\vb{p}}
  \newcommand{\vba}{\vb{A}}
\newcommand{\cl}{\textrm{cl}}
\newcommand{\eff}{\textrm{eff}}
\newcommand{\mm}{{\ensuremath{\mathsf{m}}}}
\newcommand{\nn}{{\ensuremath{\mathsf{n}}}}
\newcommand{\ts}{{\ensuremath{t_s}}}
\newcommand{\tauT}{{\ensuremath{\tau_{\scriptscriptstyle\textrm{T}}}}}
\newcommand{\tn}{{\ensuremath{t_0}}}

\newcommand{\dip}{{\ensuremath{\Delta I_{\mm\nn}}}}
\newcommand{\Xco}{\ensuremath{\textrm{X}}}
\newcommand{\Aco}{\ensuremath{\textrm{A}}}
\newcommand{\Bco}{\ensuremath{\textrm{B}}}

\newcommand{\Ilong}{I_\mathrm{long}}
\newcommand{\Itrans}{I_\mathrm{trans}}

\begin{document}

\title{Momentum transfers in correlation-assisted tunnelling}

\author{Emilio Pisanty$^{1}$}
 \email{e.pisanty11@imperial.ac.uk}
\author{Misha Ivanov$^{1,2,3}$}%
 \email{m.ivanov@imperial.ac.uk}
\affiliation{%
$^1$Blackett Laboratory, Imperial College London, South Kensington Campus, SW7 2AZ London, United Kingdom\\
$^2$Max Born Institute, Max Born Strasse 2a, 12489 Berlin, Germany\\
$^3$Department of Physics, Humboldt University, Newtonstrasse 15, 12489 Berlin, Germany
}

\date{\today}

\begin{abstract}
We consider correlation-assisted tunnel ionization of a small molecule by an intense low-frequency laser pulse. In this mechanism, the departing electron excites the state of the ion via a Coulomb interaction. We show that the wavepackets emerging from this process can have nontrivial spatial structure and give a measurable indicator of correlated multielectron dynamics during the tunnelling step. We also show that the saddle-point approximation requires special attention in this geometric analysis.
\end{abstract}
%

\maketitle


The strong correlations and interactions of electrons in close proximity are core concepts in atomic, molecular and solid-state physics. For example, in  pho\-to\-ion\-iza\-tion they feature in mechanisms like auto-ionization~\cite{FanoResonances}, giant resonances~\cite{GiantResonances}, post-ionization interaction~\cite{KnockoutShakeOff}, shake-off~\cite{KnockoutShakeOff}, shake-up~\cite{ShakeUpReview}, Auger and frustrated Auger decay~\cite{VitaliBridgette}, interatomic Coulombic decay~\cite{ICD-Averbukh-Cederbaum}, ultrafast correlation-driven hole migration~\cite{UCM-Cederbaum,HoleMigration-Kuleff}, and many others. These correlation-driven mechanisms can leave clear traces that identify them, such as the Fano line-shapes in autoionization, but the distinction between different mechanisms can also be blurry, as in the case of separating contributions of shake-up and post-ionization interaction (see e.g. Ref.~\citealp{GaugeDependenceDiagrams}).

In contrast to one-photon ionization, analyses of strong-field ionization, often viewed as optical tunnelling, have been dominated by the single active electron approximation. The inclusion of multi-electron effects beyond self-consistent field corrections~\cite{MadsenKeller} was triggered by the realization that molecular ions produced in strong laser fields are often electronically excited~\cite{ZonExcitedTunnelling,Litvinyuk-Shakeup}, and that these excitations affect all subsequent processes~\cite{Olga-ExcitedTunnelling, MultichannelSpectroscopy-Mairesse,LinSelfReview,MolecularStructureAndDynamicsHHG,AttosecondImaging}. Recent \textit{ab initio} simulations~\cite{PatchkovskiiSpanner,ExcitedIonizationInWater} and experiments~\cite{ExcitedIonizationExperiment,N2O4VibrationalExcitationExperiment} confirm that in molecules electronic 
excitations during the ionization process are a rule rather than an exception.
 
\begin{figure}[htbp]
	\centering
		\subfigure[\label{fig:1a}]{\includegraphics[scale=0.475]{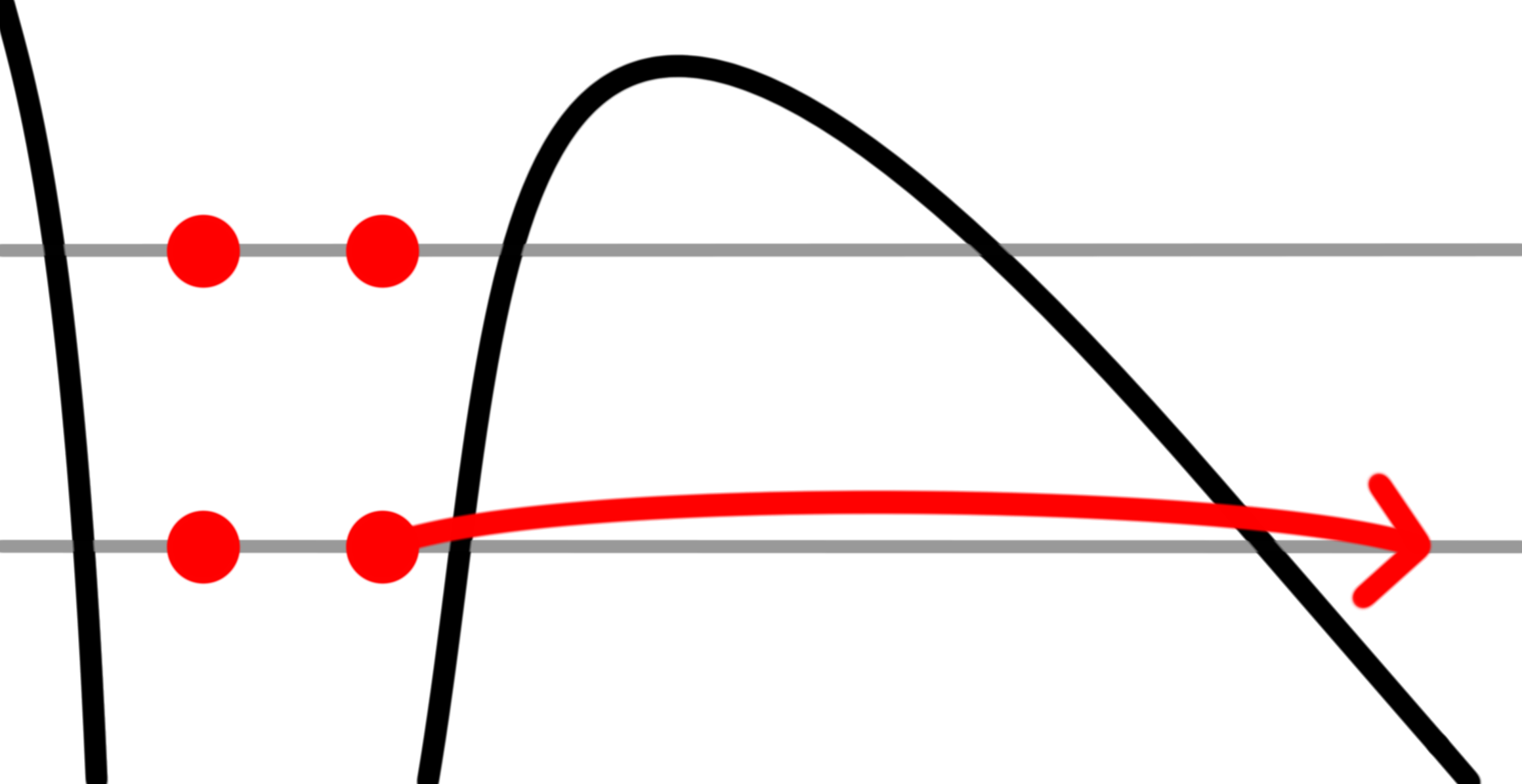}}
		\subfigure[\label{fig:1b}]{\includegraphics[scale=0.475]{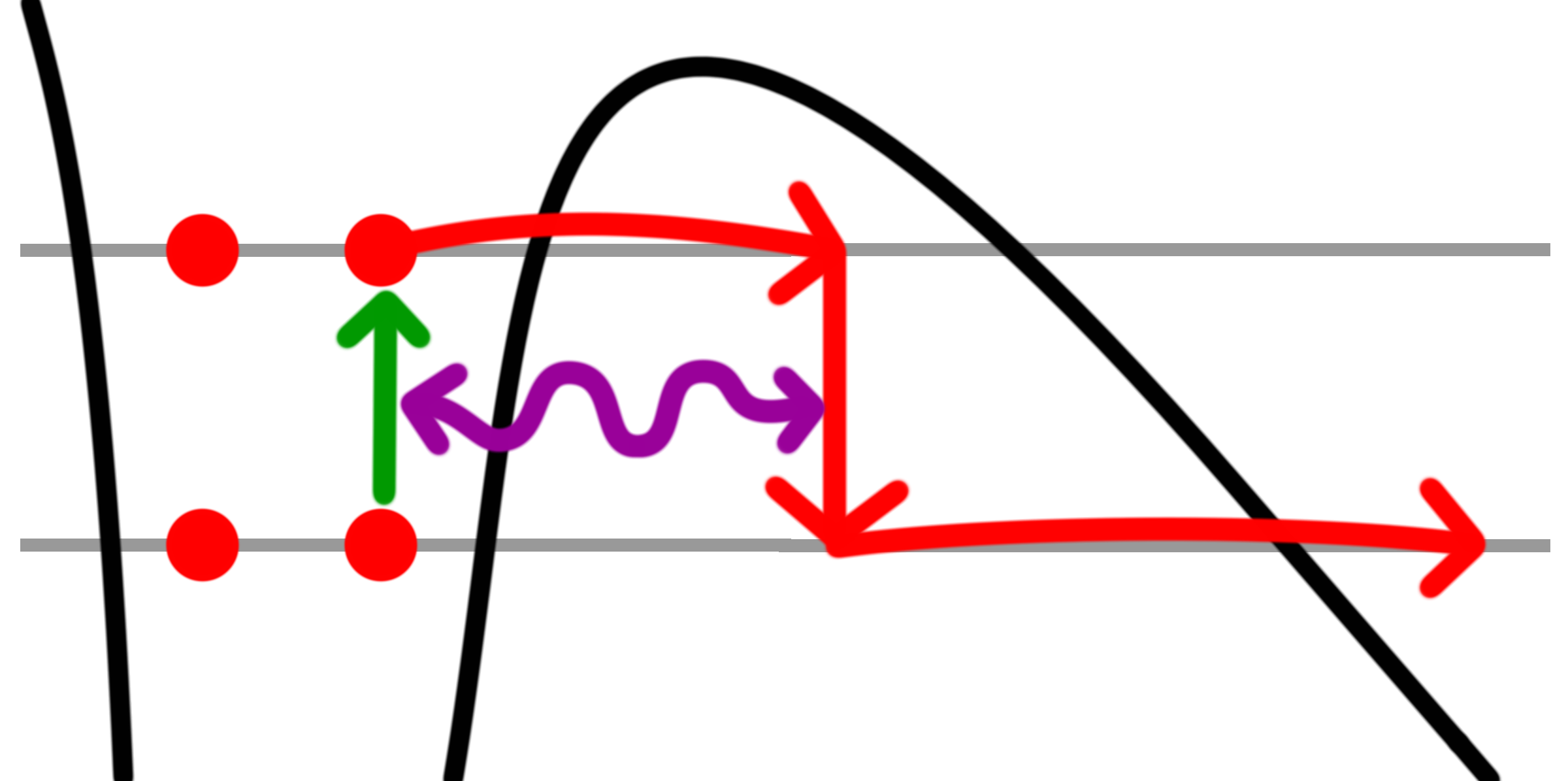}}
		\subfigure[\label{fig:1c}]{\includegraphics[scale=0.475]{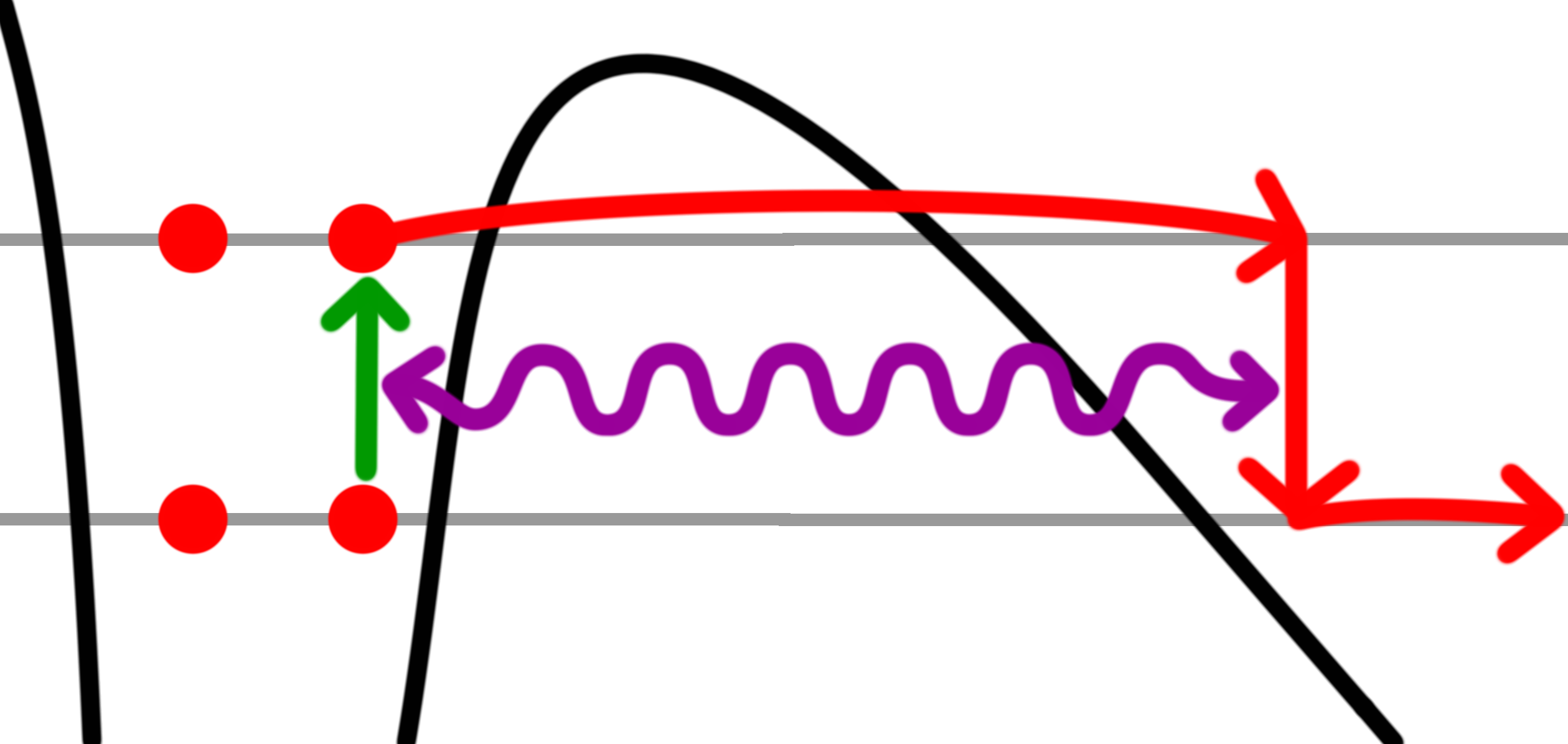}}
	\caption{Three possible ionization processes which leave the core excited: (a) the ionized electron may depart from a sub-HOMO orbital, or it may interact with the core either (b) inside the tunnelling barrier or (c) after the ionization step.
	}
	\label{fig:direct-vs-correlated}
\end{figure}

Two main mechanisms are responsible for creating an ion in an excited electronic state after optical tunnelling. First, the laser pulse may remove electron from a low-lying orbital, leaving the ionic core excited \cite{ZonExcitedTunnelling, Olga-ExcitedTunnelling,MultichannelSpectroscopy-Mairesse, LinSelfReview, MolecularStructureAndDynamicsHHG, AttosecondImaging, ZonManyBodyEffects, ZonKinetics, ZonNeon, ZonManyBody2, ZonManyElectron, GuehrScience} (shown schematically in Fig.~\ref{fig:direct-vs-correlated}(a)). Alternatively, the electron may depart from the highest occupied molecular orbital (HOMO) and subsequently excite the core through a Coulomb in\-ter\-act\-ion (shown in Figs.~\ref{fig:direct-vs-correlated}(b,c)). This can happen either inside the tunnelling barrier~\cite{OlgaMidbarrierTransitions}, shown in (b), or after the tun\-nel\-ling step \cite{Litvinyuk-Shakeup,ZonExcitedTunnelling}, shown in~(c). We refer to both (b) and (c) as correlation-assisted tunnelling.

A more formal description of these processes has recently been developed~\cite{SAEpaper,MEpaper}, which applies an analytical version of the $R$-matrix approach~\cite{RMatrix} to strong field ionization. It appears that correlation-inducing interactions (as opposed to mean-field interactions such as those studied in Ref.~\citealp{MadsenKeller}) are strong enough to influence and even dominate the ionization process. However, the calculations in Refs.~\cite{SAEpaper,MEpaper} only present total ionization rates, and these do not readily yield direct, qualitative traces of the interactions that shape the tunnelling process. 

This work looks for such traces in the angular distribution of the photoelectron. We show that correlation-ass\-is\-ted tunnelling, as shown in Figs.~\ref{fig:direct-vs-correlated}(b,c), produces wave- packets with nontrivial spatial structure. These should interfere with the direct channel to provide clear traces, detectable in angle-resolved photoelectron spectra, that multi-electron dynamics are important \textit{during} the tunnelling step. This method is independent of the choice of perturbation expansion used to obtain the diagrams in Fig.~\ref{fig:direct-vs-correlated}.

The motivation for focusing on the transverse momentum distribution is simple. If the laser field directly removes an electron from some orbital, then the outgoing wavepacket  will carry the imprints of the spatial structure of the orbital it came from~\cite{MeckelScience}. On the other hand, if the electron switches channels by inducing transitions in the ion, then the spatial structure of the outgoing wavepacket will be due to the original orbital and the nature of the ionic transition. The resulting distribution can then be different to that of the direct removal, and could therefore be used to distinguish the two contributions.

Additionally, the electron angular distribution is an important observable in its own right~\cite{MeckelScience, AngularDependenceMeasurements, AngularDependenceTheory, LinMolecularOrbitalSymmetry}, both for the information it yields directly and for its strong effect on subsequent recollision dynamics, including electron-ion diffraction and holography~\cite{ReadingDiffactionImages, YurchenkoRescattering, BlagaImaging, HuismansHolography}. Moreover, the observable coherence of the hole left in the ion during multi-channel ionization is conditioned by the overlap of the corresponding continuum electron wavepackets.

We therefore analyse in detail the angular distributions of direct ionization from orbitals below the HOMO and of the correlation-assisted contribution. The essentials of these distributions are determined by the symmetries of the orbitals and transitions involved, which then allows us to look for qualitative differences in addition to quantitative predictions. 

We include electron-electron correlations to first order in a region away from the ion (and exactly when near it)  \cite{RMatrix}. We find qualitative differences in the angular profiles of the direct and correlation-assisted electron wavepackets for dipole transitions in the ion perpendicular to the molecular axis and the polarization of the laser field. We also find important corrections to previous results~\cite{MEpaper}, which are due to a breakdown of the standard version of the saddle-point approximation in this geometry.

More specifically, we calculate the channel- and mo\-men\-tum-re\-sol\-ved ionization yield
\begin{equation}
a_\mm(\vbp) = \bra{\vbp}\!\bracket{\mm}{\Psi(T)}\text{ for }T\rightarrow\infty,
\label{eq:ionizationYieldDefinitionIntro}
\end{equation} 
where $\Psi(T)$ is the system wavefunction at large times, $\ket{\vbp}$ is a continuum state with asymptotic momentum $\vbp$, and $\ket{\mm}$ is the final state of the ion. The angle- and energy-resolved photoelectron spectrum $|a_\mm(\vbp)|^2$ should be observed in coincidence with ionic state detection on aligned molecules; such photoelectron-photoion coincidence measurements are now becoming standard for ionic states that lead to well-defined fragments~\cite{ ExcitedIonizationExperiment}. Alternative measurements could include recollision-based indirect imaging schemes such as two-dimensional high-harmonic generation spectroscopy~\cite{ResolvingExitTimes} and laser-induced electron ho\-lo\-gra\-phy~\cite{ReadingDiffactionImages,HuismansHolography} and dif\-fract\-ion 
\cite{ReadingDiffactionImages,YurchenkoRescattering,BlagaImaging, LinSelfReview}, which are all intrinsically sensitive to the ionic state. We concentrate on the so-called direct electrons, which do not rescatter with the core; these dominate the photo-electron spectrum up to energies of $2U_p$~\cite{directElectrons}, and their momentum distribution can now be measured and characterized with high accuracy~\cite{ArissianCorkumPRL}.

\begin{figure}[htbp]
	\centering
		\includegraphics[width=8.5cm]{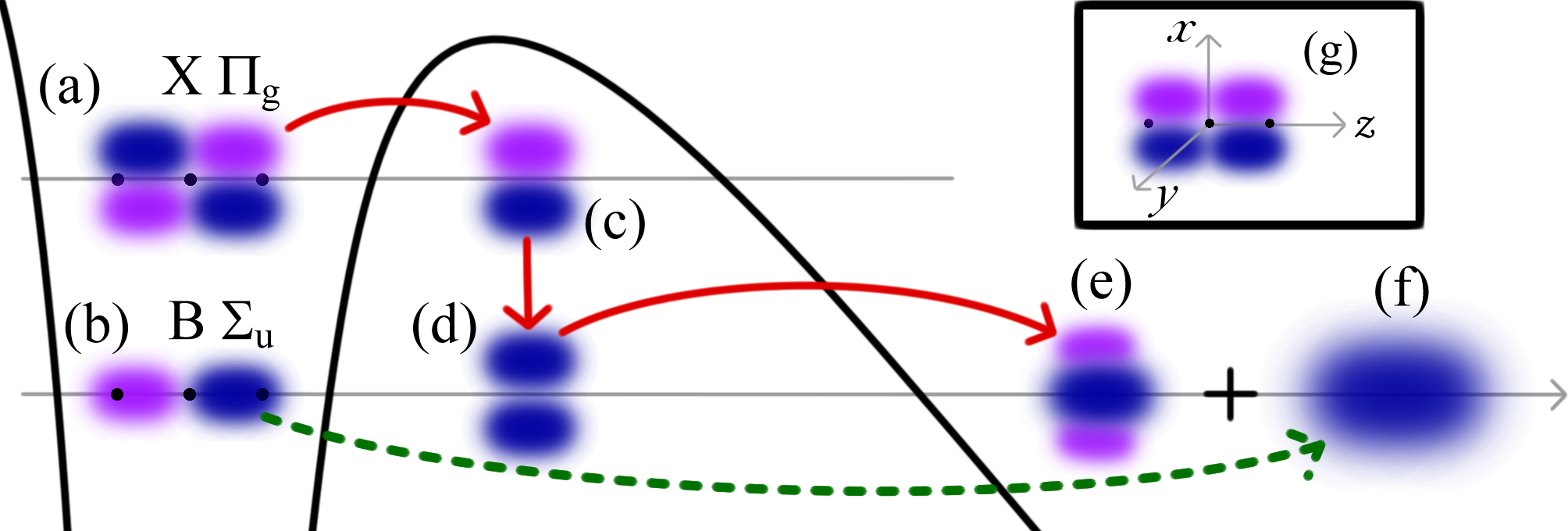}
	\caption{Correlation-assisted ionization of CO$_2$. An electron can (a) ionize from HOMO and (b) change to an excited channel in (c) a mid-barrier transition. This subjects it (c) to the dipole potential of the transition charge, (g), which changes the relative phase of the two lobes (d). This double-slit wavefunction then diffracts to multiple lobes (e). This contrasts with direct ionization on the excited $\Bco$ channel, which has a single lobe (f).}
	\label{fig:CO2-X-to-B-coupling}
\end{figure}

To illustrate the physical origin of our results, consider the tunnel ionization of CO$_2$ with the laser polarization along the internuclear axis, as shown in Fig.~\ref{fig:CO2-X-to-B-coupling}. The leading perpendicular transition is from the ground-state channel of $\textrm{CO}_2^+$, $\Xco\, \Pi_\text{g}$, to its second excited channel, $\Bco\, \Sigma_\text{u}$. These correspond to removal of an electron from HOMO and from HOMO-2, respectively, which are shown in Figs.~\ref{fig:CO2-X-to-B-coupling}(a) and \ref{fig:CO2-X-to-B-coupling}(b). 

Here HOMO has a nodal plane along the laser polarization, with two lobes of opposite phase, and the outgoing wavepacket inherits this structure. This gives a similar configuration in momentum space, where lobes now represent counter-propagating waves in position space that destructively interfere at the nodal plane. (Such structures have been observed in experiment~\cite{MeckelScience}.)

At the moment of the correlation interaction (which is later integrated over), this wave\-pack\-et is impulsively subjected to the correlation potential. For our perpendicular transition the correlation potential is essentially given by $d_{\Bco\Xco}{x}/{z^3}$, with axes as in Fig.~\ref{fig:CO2-X-to-B-coupling}(g). (We note that there is an analogous channel along the $y$ axis, coming from the degenerate ground state, which will restore cylindrical symmetry to the final result.) This is linear in the transverse coordinate $x$, so the force is constant. In momentum space, this operator is then proportional to the derivative $\frac{\partial}{\partial k_x\!\!\!{}}\,$, and transforms the two-lobed momentum wavefunction into a three-lobed one as shown in Fig.~\ref{fig:CO2-X-to-B-coupling}(e). 

The physical picture is most clearly cast in terms of angular momentum. The $\Xco$ channel is a $\Pi$ state, which means that the outgoing electron and the hole in the core both have angular momenta $L=\pm 1$ about the laser polarization, in opposite directions. The $\Bco$ channel, on the other hand, is a $\Sigma$ state with zero angular momentum in the core. Inducing an $\Xco\rightarrow\Bco$ transition thus requires the outgoing electron to `wind down' the core, returning its angular momentum through the reaction force. This exchange of transverse momentum creates the central lobe.

The lateral lobes in the final momentum distribution are interference effects coming from the interaction region. In position space, the initial tunnelling wavepacket is Gaussian in the transverse direction~\cite{AnatomyOfStrongFieldIonization} with a node of the form $\psi\propto x e^{-\frac{1}{2\tau}x^2}$. The impulsive application of the dipole potential transforms it to the form $\psi'\propto x^2 e^{-\frac{1}{2\tau}x^2}$; the final momentum distribution is the Fourier transform of this wavefunction. The situation is then essentially interference from a double slit (Fig.~\ref{fig:CO2-X-to-B-coupling}(d)) with three of the fringes visible.

We now proceed to the formal analysis of this mechanism. Correlation interactions are weak away from the ion, where the interactions are favoured by the smaller size of the total barrier. The physical amplitude of ionization $a_\mm(\vbp)$ can therefore be expanded in a time-dependent perturbation expansion in the correlation interaction potential,
\begin{equation}
\hat{V}_{ee}^\nn=\sum_i \frac{1}{|\hat\vbr-\hat\vbr_i|}-\bra{\nn}{\sum_i \frac{1}{|\hat\vbr-\hat\vbr_i|}}\ket{\nn},
\label{eq:CorrelationPotential}
\end{equation}
in which the zeroth-order terms are the usual mean-field interactions, and the correlation-assisted ionization signal is the first-order correction. (Here $\nn$ is the initial ionic state, and describes the entrance channel.) This contribution can be written in the form
\begin{align}
a_\mm^{(1)}(\vbp) 
& =
-i
\sum_\nn
\int \!\!\d t \!
\int\!\!\d\vbk   \,
e^{-\frac i2 \int_{t}^T(\vbp+\vba(\tau))^2\d\tau}
\nonumber\\ & \qquad\quad\times
\bra{\vbp}\! \bra{\mm}{\hat{V}_{ee}^\nn}\ket{\nn}\!  \ket{\vbk} 
e^{i\dip t}
\nonumber\\ & \qquad\quad\times
R_\nn(\vbk) 
e^{-\frac i2 \int_{\ts}^{t}(\vbk+\vba(\tau))^2\d\tau} 
e^{i I_{p,\nn} \ts}.
\label{eq:SchematicTimeIntegral}
\end{align}

This expression is derived formally in Ref.~\citealp{MEpaper} using the analytical $R$-matrix (ARM) theory, and it can be understood intuitively as follows. The electron is ionized at a complex time $\ts=\tn + i \tauT$ to the entrance channel $\nn$ with momentum $\vbk$, which has an amplitude form factor $R_\nn(\vbk)$ and an exponential tunnelling penalty of $|e^{i I_{p,\nn} \ts}|=e^{- I_{p,\nn} \tauT}$, where $I_{p,\nn}$ is the ionization potential on channel~$\nn$. (We use atomic units throughout.)

The electron then propagates through complex time, which introduces a damping factor $e^{-\frac i2 \int_{\ts}^{t}(\vbk+\vba(\tau))^2\d\tau}$, until the interaction time $t$, at which it is multiplied by the correlation interaction potential $V_{ee}^\nn(\hat\vbr)$. It then propagates until its detection with momentum $\vbp$ on channel $\mm$, which introduces a further factor of $e^{-\frac i2 \int_{t}^T(\vbp+\vba(\tau))^2\d\tau}$ and a further damping of $e^{i\dip t}$, where $\dip=I_{p,\mm}-I_{p,\nn}$ is the increase in the height of the tunnelling barrier. 

The amplitude in Eq.~\eqref{eq:SchematicTimeIntegral} arises from a multiconfiguration wavefunction expansion of the form
\begin{equation}
\ket{\Psi^N}=\hat{A}\sum_j\ket{\chi_j}\ket{\Phi_j^{N-1}},
\label{eq:MCWF-expansion}
\end{equation}
where the $N$-electron state $\ket{\Psi^N}$ of the full system is expressed, when the photoelectron is outside a suitably large spherical boundary, in terms of the exact ionic eigenstates $\ket{\Phi_j^{N-1}}$ and the corresponding photoelectron wavepackets $\ket{\chi_j}$, suitably antisymmetrized. The photoelectron wavepackets are `launched' into the continuum by the Dyson orbitals $\ket{\psi_j^\text{Dys}}=\sqrt{N}\bracket{\Phi_j^{N-1}}{\Psi_g^N}$ (which are the one-electron orbitals pictorially shown in Fig.~\ref{fig:CO2-X-to-B-coupling} and colloquially referred to above), using a Bloch operator to ensure continuity of the wavefunction at the boundary, as is standard in the $R$-matrix approach. This boundary is chosen deep enough into the barrier that exchange (and therefore the antisymmetrization in \eqref{eq:MCWF-expansion}) can be neglected, which is the crucial simplifying assumption brought in by the (standard) $R$-matrix method \cite{RMatrix}.  

\begin{figure}[htbp]
\begin{center}
    	%
	    %
      \includegraphics[width=6.8cm]{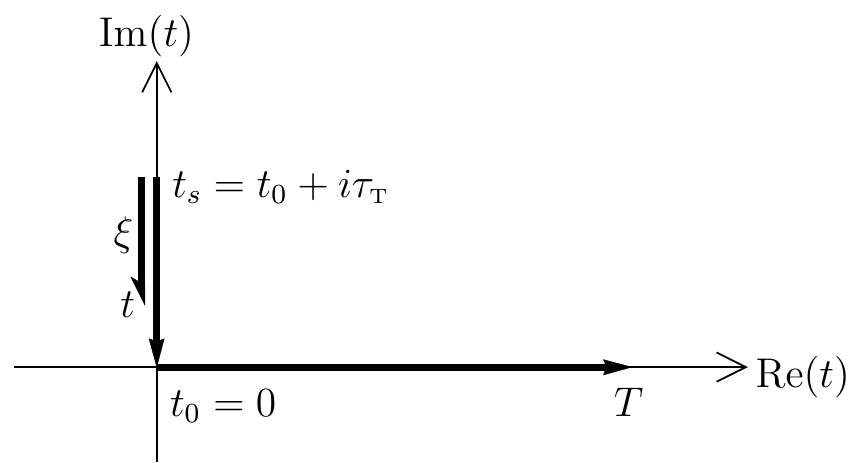}
	\caption{Contour for the complex integration over the interaction time $t$, as described in the text.}
	\label{fig:integrationcontour}
\end{center}
\end{figure}

We consider a linearly polarized sinusoidal field with vector potential $\vba(t)=-\frac F\omega\hat{\vb{e}}_z \sin(\omega t)$. The initial channel and momentum, $\nn$ and $\vbk$, are undetermined and must be coherently summed over. The one-electron form factor $R_\nn(\vbk)$ comes from a temporal saddle-point analysis of the ionization step, which yields a quantum-orbit picture of ionization \cite{QuantumOrbits,MEpaper}. Here the electron leaves the atom at a complex ionization time $\ts=\tn + i \tauT$  which obeys the equation
\begin{equation}
\frac12 \left(\vbp+\vba(\ts)\right)^2 +I_{p,\nn}=0.
\label{eq:SaddlePointEquation}
\end{equation}
The interaction time $t$ is integrated over on a contour that starts at $\ts$, drops to its real part, $\tn$, on the real time axis, and continues along the real axis until the measurement time $T$, as shown in Fig.~\ref{fig:integrationcontour}. We concentrate for simplicity on the contribution of the part of this contour moving parallel to the imaginary time axis, which represents interactions before the barrier exit. This is the dominant contribution: the electron leaves the barrier with a constant acceleration and interacts with a dipolar field, so the time integral outside the barrier is of the form $\int t^{-5} \d t$ and converges rapidly. Similarly, the exponential factor $e^{i\dip t}$ forces most of the contributions to come from near the tunnel exit, where the total barrier is the thinnest; this is far enough from the ion that a perturbation expansion over electron correlation is justified.

In general, time, and therefore the classical trajectory 
\begin{equation}
\vbr_\cl(t)=\int_\ts^t(\vbp+\vba(\tau))\d\tau,
\label{eq:ClassicalTrajectory}
\end{equation} 
are complex-valued. For simplicity, we consider ionization at a peak of the field, with $\Re(\ts)=\tn=0$, which corresponds to the peak of the parallel momentum distribution at $p_z=0$, the maximum ionization amplitude, and a real-valued classical trajectory
\begin{equation}
z_\cl(t)=\frac{F}{\omega^2}\left(\cosh(\omega \Im(t))-\cosh(\omega\tauT)\right).
\label{eq:RealZcl}
\end{equation} 

We focus on a single perpendicular  transition, such as the \Xco--\Bco transition in CO$_2^+$. We also ignore the laser-induced polarization of both states, as the two states do not couple in this geometry, and mixing with other states does not affect the nodal geometry which is crucial for our results. We use Volkov wavefunctions~\cite{VolkovWavefunctions} for the initial and final states of the continuum electron, though eikonal corrections~\cite{EVApaper}, which include the Coulomb field of the ion, can be included for better accuracy.

Correlation- induced transitions are favoured closer to the barrier exit, since this minimizes the exponential penalty on tunnelling through a thicker barrier after the transition~\cite{OlgaMidbarrierTransitions}, as shown in Fig.~\ref{fig:CO2-X-to-B-coupling}. Therefore, we model the interaction potential as a softened dipole in the approximation of small tunnelling angles, with
\begin{equation}
\bra{\vbr'}\!\bra{\mm}{\hat{V}_{ee}^{\nn}}\ket{\nn}\!\ket{\vbr}=\frac{d_{\mm\nn}x}{\left(z^2+\sigma^2\right)^{3/2}} \bracket{\vbr'}{\vbr}\!,
\label{eq:SoftenedDipole}
\end{equation} 
though more accurate models exist which are also amenable to analytical integration and give similar results \cite{MResReport}. The choice of a model for the potential also depends on its behaviour under analytic continuation (softened dipoles, for example, have unphysical poles at $z=\pm i \sigma$), but this is not an issue at the peak of the field since the classical trajectory, expression \eqref{eq:RealZcl}, is real.

For small tunnelling angles, the single-electron form factor is similar to that for tunnelling from a hydrogenic $P_x$ orbital for the initial state, 
\begin{equation}
R_\nn(\vbk)=C_\nn(k_z) k_x=\frac{C_{0,\nn}}{\sqrt{iS_V''(\ts)}}k_x
\label{eq:FormFactor}
\end{equation}
where $S_V(t)=\frac i2\int_T^t(\vbp+\vba(\tau))^2\d\tau+i I_{p,\nn} t$ is the Volkov action. For our purposes, this factor simply embodies the nodal structure of the outgoing wavepacket.

Under these conditions, the correlation-driven yield separates into a product of longitudinal and transverse parts~as
\begin{subequations}
\label{IlongItransDef}

{\allowdisplaybreaks \smaller
\begin{align}
a_\mm^{(1)} (\vbp) 
& =-ie^{-\frac i2 \int_{\tn}^T(\vbp+\vba(\tau))^2\d\tau}
\nonumber \\ & \phantom{=}\times
\int_\ts^\tn \d t \,e^{i I_{p,\nn} \ts}\,e^{i\dip t}\,\Ilong\times\Itrans
,
\label{eq:YieldSeparation}
 \\
\textrm{for }
\Ilong
 &=
e^{
      -\frac i2 \int_{t}^{\tn}(p_z+A_z(\tau))^2\d\tau
  }
\label{eq:IlongDef}
\int
\frac{
  \d z  \d k_z 
	}{
	2\pi
}
\frac{
  C_\nn(k_z)
e^{i(k_z-p_z)z}
	}{
	\left(z^2+\sigma^2\right)^{3/2}
}
\nonumber \\ & \phantom{=}\times
e^{-\frac i2 \int_{\ts}^{t}(k_z+A_z(\tau))^2\d\tau} 
\\
\textrm{and }
\Itrans
 &=
e^{\frac {\xi-\tauT}{2}p_\perp^2}
\int
\frac{
  \d^2\vbr_\perp
	\d^2\vbk_\perp 
	}{
	(2\pi)^2
}
\,
d_{\mm\nn}x
\,
k_x
\label{eq:ItransDef}
e^{i(\vbk_\perp-\vbp_\perp)\cdot\vbr_\perp}
e^{-\frac{\xi}{2}k_\perp^2}
,
\end{align}
}
\end{subequations}
where $\xi=i(t-\ts)\geq0$, as shown in Fig.~\ref{fig:integrationcontour}. The longitudinal integral can be approximated using saddle-point methods, which additionally allows for more general pulse shapes, while the transverse integral is simple enough to handle.

The integrals over $y$, $k_y$ and $k_x$ are simple and the saddle-point method is exact, yielding
\begin{equation}
\Itrans
 =
\frac{
e^{-\frac{\tauT}{2}p_\perp^2}
	}{
	\sqrt{2\pi\xi}
}
\int \!  \d x
\,
d_{\mm\nn}x
\,
\frac{ix}{\xi}
e^{-\frac{1}{2\xi}(x+i \xi p_x)^2}.
\label{eq:ItransBeforexIntegral}
\end{equation}
Here the factor $ix/\xi$ is the saddle point on the $k_x$ integral and has the interpretation of the momentum that will advance the electron by $x$ in time $-i\xi$.

The integral over $x$ can now be performed to give
\begin{equation}
\Itrans
 =
i d_{\mm\nn}(1-\xi \,p_x^2)
e^{-\frac{\tauT}{2}p_\perp^2}.
\label{eq:ItransAfterxIntegral}
\end{equation}
This already gives the transverse momentum profile we have been looking for. Its general characteristics will not be altered by the temporal integration over $\xi $.

To obtain a final expression, we apply the saddle-point approximation to the longitudinal integral, which gives
\begin{equation}
\Ilong
=
\frac{
  C_\nn(p_z)
  e^{-\frac i2 \int_{\ts}^{\tn}(p_z+A_z(\tau))^2\d\tau} 
	}{
	\left(z_\cl^2(t)+\sigma^2\right)^{3/2}
}
,
\label{eq:ItransSolution}
\end{equation}
and therefore an ionization yield of
\begin{align}
a_\mm^{(1)}(\vbp)
& =
- 
e^{i I_{p,\nn}\ts}
e^{-\frac i2 \int_{\ts}^{T}(\vbp+\vba(\tau))^2\d\tau} 
  C_\nn(p_z)
\nonumber \\ &  \times
\int_0^\tauT
\frac{
  i d_{\mm\nn}(1-\xi \,p_x^2)
	}{
	\left(z_\cl^2(\ts-i\xi)+\sigma^2\right)^{3/2}
}
e^{\dip (\xi-\tauT)}
\d  \xi
.
\label{eq:FinalIonizationYield}
\end{align}

\begin{figure}[htbp]
\begin{center}
	\includegraphics[width=8.5cm]{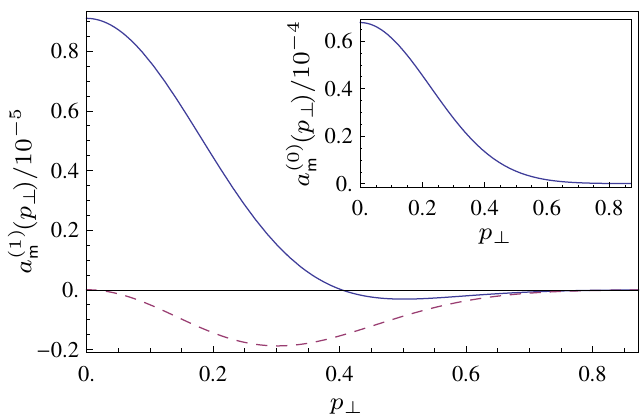}
  \caption{
  Momentum-resolved ionization yield for the $\Xco\rightarrow\Bco$ perpendicular dipole transition in parallel-aligned CO$_2$, with the corresponding direct amplitude in the inset. The dashed curve is the prediction from the spatial saddle-point approximation. The parameters used are $F=0.05$ and $\omega=0.055$ (so $I\approx 9 \times 10^{13}\,\textrm{W}/\textrm{cm}^2$ and $\lambda=800\,\textrm{nm}$), $d_{\mm\nn}=0.175$, $\sigma=2.19$, $C_{0,\Xco}=-0.23$, $C_{0,\Bco}=0.18 i$, $I_{p,\Xco}=0.5064$, and  $I_{p,\Bco}=0.6644$. (All quantities are in atomic units unless noted). 
	}
	\label{fig:graphs-CO2-X-B}
\end{center}
\end{figure}

Before analysing this expression in depth, we remark that the standard saddle-point approximation fails when applied to Eq.~\eqref{eq:ItransBeforexIntegral} en route to Eq.~\eqref{eq:ItransAfterxIntegral}. Indeed, it returns  the term in $-\xi p_x^2$, which is only a good approximation when $\xi p_x^2\gg 1$ and is shown as a dashed line in Fig.~\ref{fig:graphs-CO2-X-B}. However, there will always be small enough momenta (giving on-axis ionization) for which this fails. The saddle point at the classical trajectory $x_\cl=-i\xi p_x$ is then close to a zero of the prefactor $x^2$ in Eq.~\eqref{eq:ItransBeforexIntegral}, which can no longer be considered slow. This causes Ref.~\citealp{MEpaper} to underestimate correlation-assisted tunnelling in this geometry.

The accuracy of the saddle point calculation can be restored with the use of second-order terms~\cite{BruijnAsymptotics,GerlachSPAonline}:
\begin{align}
\label{eq:SPASecondOrderTerms}
&
\int_A^B F(\zeta)e^{\rho \varphi(\zeta)}\d\zeta
=
\sqrt{\frac{2\pi}{\rho}} 
\frac{e^{\rho\phi(\zeta_0)}}{\left[-\phi''(\zeta_0)\right]^{1/2}}
\times
\\ & \qquad\quad\times
\left[F(\zeta_0)
-\frac{F''(\zeta_0)}{2\rho\varphi''(\zeta_0)}
+\frac{1}{2!}\frac{F^{(4)}(\zeta_0)}{\left(2\rho\varphi''(\zeta_0)\right)^2}
-\cdots\right]
.
\nonumber
\end{align}
In general, using as many derivatives of the prefactor as the order of the highest expected zero is sufficient. More practically, a zero answer should be distrusted unless there is a specific reason (such as angular momentum conservation) that dictates it. Other failures of the standard saddle-point approximation in strong-field phenomena have been reported previously~\cite{LeinSPAFailure, InterferenceSPAfailure, UniformTemporalSPA}, which underlines the need for a careful evaluation of such approximations when applied in new settings.

To extract the radial profiles from expression~\eqref{eq:FinalIonizationYield} only the 
temporal integration over $t=\ts-i\xi$ remains. This cannot be done analytically due to the time dependence of $z_\cl(t)$, which is given by Eqs.~\eqref{eq:ClassicalTrajectory} and \eqref{eq:RealZcl}, but since the dependence on $p_\perp$ (through the time $\ts$) is weak, the angular profile remains of the form 
\begin{equation}
a_\mm^{(1)}(\vbp)
\propto
(1-\tau_\eff\, p_x^2)e^{-\frac \tauT 2 p_\perp^2}
\textrm{ with }\tau_\eff\sim\tauT
,
\label{eq:EssentialAngularProfile}
\end{equation}
with three distinct lobes of opposite phase. 
It is shown in Fig.~\ref{fig:graphs-CO2-X-B}, corresponding to the contribution from a single half-cycle of the laser field. As  mentioned above, the three lobes have a direct interpretation as interference fringes from a mid-barrier double-slit wavefunction. (Upon inclusion of the $\Pi_y$ ground state, on the other hand, the outer lobes will become a ring, restoring the rotational symmetry.)

This angular distribution is in contrast with the one for the direct channel from a $\Sigma$ state~\cite[][eq. (75)]{SAEpaper}, which goes as
\begin{equation}
a_\mm^{(0)}(\vbp)=
e^{i I_{p,\mm} \ts}R_\mm(\vbp)e^{-\frac i2 \int_\ts^T\left(\vbp+\vba(\tau)\right)^2\d\tau}
\propto
e^{-\frac{\tauT}{2}p_\perp^2}
,
\label{eq:SAEyield}
\end{equation}
with a single lobe, and is shown inset in Fig~\ref{fig:graphs-CO2-X-B}. The two contributions are of comparable magnitude, which was found in Ref.~\citealp{MEpaper} for other alignment angles.

There is, then, considerable structure in the correlation-assisted wavefunction, as opposed to the structureless gaussian that is generally predicted by single-active-electron theories \cite[and references within]{ArissianCorkumPRL}. Since the final ionic state is the same, both wavepackets will add coherently. The signals are comparable -- with transitions to the first excited state, $\Aco\,\Pi_u$, even stronger due to a reduced $\dip$ -- so there should be considerable interference and thus nontrivial structure in the final amplitude. 

\begin{figure}[htbp]
\begin{center}
    \includegraphics[width=8.5cm]{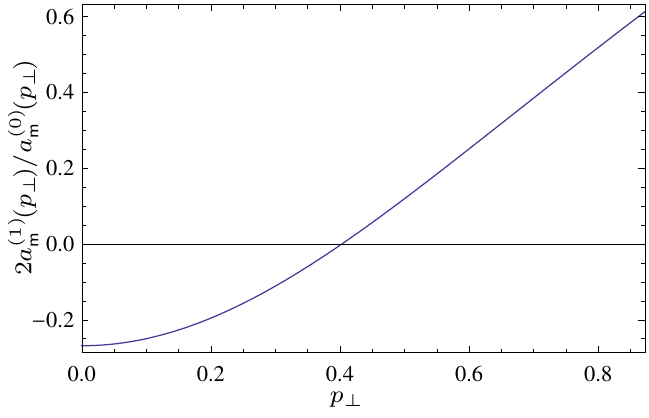}
    \caption{
       Ratio between the correlation-assisted and the direct signals for the $\Xco\rightarrow\Bco$ perpendicular dipole transition in parallel-aligned CO$_2$, with the same parameters as Fig.~\ref{fig:graphs-CO2-X-B}. This contributes to the total amplitude in a heterodyne-like scheme, as in Eq.~\eqref{eq:channelsratio}.
	}
	\label{fig:channelsRatio}
\end{center}
\end{figure}

This interference between both channels will depend, in particular, on their relative phase, which is nontrivial as the trajectory $z_\cl(t)$ is imaginary for ionization times after the peak of the field, and contributes a phase to Eq.~\eqref{eq:FinalIonizationYield}. In particular, we expect that the interference between direct and correlation-assisted tunnelling will change across the electron's longitudinal momentum distribution. Thus, the simplified model we present here cannot provide an accurate prediction of the final structure. However, an order-of-magnitude estimate is indeed possible: the total detection probability will be, approximately, 
\begin{align}
|a_\mm(\vbp)|^2
&=|a_\mm^{(0)}(\vbp)+a_\mm^{(1)}(\vbp)|^2 \nonumber
\\& \approx |a_\mm^{(0)}(\vbp)|^2(1+2a_\mm^{(1)}(\vbp)/a_\mm^{(0)}(\vbp)),
\label{eq:channelsratio}
\end{align}
so that deviations from gaussianity should be observed at the level of $2a_\mm^{(1)}(\vbp)/a_\mm^{(0)}(\vbp)$, which is shown in Fig.~\ref{fig:channelsRatio}. For the case of $\Xco\rightarrow\Bco$ transitions in CO$_2$, this can be as great as 20\%.

Experimental detection of nontrivial structure in these geometries would imply multielectron dynamics took place during the tunnelling step. Additionally, if the details of the inter-channel interference depend on controllable parameters, it would open the door to direct shaping of the electron's wavefunction at the tunnel exit.

This work was funded by EPSRC Program Grant \textsc{EP/ I032517/\-1} and the CORINF Training Network; EP gratefully acknowledges support from CONACYT. We thank Lisa Torlina, Olga Smirnova and Serguei Patchkovskii for essential collaboration.



\end{document}